\documentclass[aps,prd,nofootinbib,superscriptaddress,preprintnumbers,twocolumn]{revtex4-1}
\usepackage[latin9]{inputenc}
\usepackage{bm}
\usepackage{amsmath}
\usepackage{amsthm}
\usepackage{amssymb}

\begin{document}

\preprint{YITP-18-85, IPMU18-0128}

\title{Black holes and stars in the minimal theory of massive gravity}

\author{Antonio De Felice}
\email{antonio.defelice@yukawa.kyoto-u.ac.jp}
\affiliation{Center for Gravitational Physics, Yukawa Institute for Theoretical Physics, Kyoto University, 606-8502, Kyoto, Japan}
\author{Fran\c{c}ois Larrouturou}
\email{francois.larrouturou@ens.fr}
\affiliation{D\'{e}partement de physique de l'ENS, \'{E}cole Normale Sup\'{e}rieure, CNRS, PSL Research University, 75005 Paris, France}
\affiliation{Center for Gravitational Physics, Yukawa Institute for Theoretical Physics, Kyoto University, 606-8502, Kyoto, Japan}
\author{Shinji Mukohyama}
\email{shinji.mukohyama@yukawa.kyoto-u.ac.jp}
\affiliation{Center for Gravitational Physics, Yukawa Institute for Theoretical Physics, Kyoto University, 606-8502, Kyoto, Japan}
\affiliation{Kavli Institute for the Physics and Mathematics of the Universe (WPI), The University of Tokyo Institutes for Advanced Study, The University of Tokyo, Kashiwa, Chiba 277-8583, Japan}
\author{Michele Oliosi}
\email{michele.oliosi@yukawa.kyoto-u.ac.jp}
\affiliation{Center for Gravitational Physics, Yukawa Institute for Theoretical Physics, Kyoto University, 606-8502, Kyoto, Japan}

\date{\today}

\begin{abstract}
In this letter, we show that any solution of general relativity (GR) that can be rendered spatially flat by a coordinate change is also a solution of the self-accelerating branch of the minimal theory of massive gravity (MTMG), with or without matter. We then for the first time obtain black hole and star solutions in a theory of massive gravity that agree with the corresponding solutions in GR and that are free from strong coupling issues. This in particular implies that the parametrized post-Newtonian parameters $\beta^{\rm PPN}$ and $\gamma^{\rm PPN}$ are unity, as in GR. We further show how these solutions can be embedded in a cosmological setting. While cosmological scales have already been considered in previous works, this is the first study of the phenomenology at shorter scales of the self-accelerating branch of MTMG.
\end{abstract}

\maketitle

\noindent{\bf\em Introduction.}
With the first observation of gravitational waves (GW) from a binary black hole merger \cite{GW150914}, there is no doubt that black holes do exist in our Universe. Black hole space-times have been central to research in gravity since the discovery of the Schwarzschild solution~\cite{Schw_OP} in the context of general relativity (GR). One of the strongest reasons for this is their simplicity, allowing one to describe celestial objects with purely analytic tools. The Schwarzschild solution can also be used as a part of the description of systems including matter: the space-time outside quasi-spherical objects such as stars and planets (without rotation) is well approximated by this solution. The exterior vacuum solution can then be connected to an interior solution with matter. In GR, the existence of black holes and, more generally, of spherically symmetric configurations has proven to be a theoretical asset as much as it is of course a phenomenological necessity. 

In the latest decades, the search for theories going beyond general relativity has grown, and it has been important to test these theories in multiple ways. On astrophysical scales, one may constrain a theory of gravity by different means; a well-known example is the use of parametrized post Newtonian (PPN) constraints \cite{Will}. Our present work is related to not only this but also another avenue: showing the existence of black hole and star solutions without strong coupling, and checking their precise phenomenology. Black holes and stars are ubiquitous and, as such, unavoidable elements that every theory of gravity should be able to describe. 

Among modified theories of gravity, massive gravity is an archetype of infrared (IR) modification, first formalized by Fierz and Pauli \cite{Pauli-Fierz}. Then, the first theory to successfully eradicate the Boulware-Deser (BD) ghost, previously thought to appear generically in massive gravity \cite{BDghost}, was de Rham-Gabadadze-Tolley (dRGT) massive gravity \cite{dRGT}. This theory was also shown to be unique under some assumptions, in particular Lorentz invariance and the restriction to the sole metric field for the gravitational sector. However, because of the lack of stable Friedmann-Lema\^{i}tre-Robertson-Walker (FLRW) cosmologies in dRGT massive gravity \cite{DeFelice:2012mx}, theories beyond dRGT have been developed (see e.g.\ \cite{bigravity,quasidilaton,massvarying,chameleonbigravity,mqd}). One of those, called the minimal theory of massive gravity (MTMG) \cite{mtmg,mtmg_pheno}, a Lorentz symmetry violating theory, has the special feature that it only propagates two tensor modes instead of the five present in dRGT theory. These modes are massive but travel at the speed of light in the subhorizon limit, thus passing with flying colors the recent constraints \cite{BNS}, while the graviton mass term contributes to the late-time acceleration of the Universe. 

In the context of black hole solutions, massive gravity has faced challenges. In dRGT massive gravity, two branches of static solutions are present (see e.g.\ \cite{BHinmg}). One of them, in which both metrics are proportional to each other, is unstable under both radial perturbations and superradiant instability \cite{BHinmg_bidiag}, and furthermore possesses coordinate-invariant singularities at the horizon \cite{BHinmg_singularity}.
While the second is claimed to be stable both in the radial and modal senses \cite{BHinmg_nonbidiag} and non-singular (anti-)de Sitter Schwarzschild solutions can be found, it suffers from infinitely strong coupling \cite{BHinmg_strongcoupling}. Note that the aforementioned problems can be bypassed by including a time dependence in the description \cite{BHinmg_strongcoupling, RosenBH} and thus deviation from corresponding solutions in GR. 
For other theories of massive gravity, such as MTMG, these issues have not yet been explored as thoroughly. It is thus interesting to determine whether MTMG faces the same problems, or, conversely, if healthy static solutions, such as black holes, can be found.

In this work, we answer positively to the latter question, and then extend our answer to time-dependent solutions such as collapsing matter. We first present briefly the theory and show the correspondence between GR and MTMG solutions. Then, as corollaries of this result, the black hole and star solutions are explicitly derived, their matching with cosmology being also discussed.

\noindent{\bf\em MTMG in a nutshell.}
MTMG is a theory of massive gravity that propagates only the two tensor modes. One may write it as a precursor theory that breaks Lorentz invariance, together with adequate constraints removing the extra degrees of freedom. MTMG thus contains, in the so-called unitary gauge, a dynamical metric $g_{\mu\nu}$ and a fiducial metric $f_{\mu\nu}$ as well as Lagrange multipliers $\lambda$ and $\lambda^i$ ($i \in \{1,2,3\}$). 

The metric formulation of MTMG relies on an Arnowitt-Deser-Misner (ADM) foliation of space-time, given by
\begin{equation}
 \mathrm{d}s^2
  =  - N^2 \mathrm{d} t^2
  + \gamma_{ij}(\mathrm{d} x^i + N^i \mathrm{d} t)(\mathrm{d} x^j + N^j \mathrm{d} t)\,,
\end{equation}
where $\mathrm{d}s^2 \equiv g_{\mu\nu} \mathrm{d}x^{\mu} \mathrm{d}x^{\nu}$, and an equivalent decomposition using tilded variables for the fiducial sector. The three-dimensional fiducial metric $\tilde{\gamma}_{ij}$ enters the theory only via the combinations $\mathfrak{K}^i_{\ j} \equiv \gamma^{ik}\tilde{\gamma}_{kj}$ and $\tilde{\zeta}^i{}_j \equiv \tilde{\gamma}^{ik}\,\partial_t\tilde{\gamma}_{kj}/(2\tilde{N})$, where $\gamma^{ij}$ and $\tilde{\gamma}^{ij}$ are inverses of $\gamma_{ij}$ and $\tilde{\gamma}_{ij}$, respectively. The action for MTMG is then given by
\begin{equation}
    S_{\rm MTMG} = S_{\rm GR} + S_{\rm mat} -\frac{M_{\rm Pl}^2 m_g^2}{2} \int \!\mathrm{d}^4x \sqrt{-g} \, \mathcal{W},
\end{equation}
where $S_{\rm GR} = (M_{\rm Pl}^2/2) \int \mathrm{d}^4x \sqrt{-g} R$ is the 
Einstein-Hilbert action, $S_{\rm mat}$ is the action for matter minimally coupled to the metric $g_{\mu\nu}$, $M_{\rm Pl}$ and $m_g$ are respectively the Planck scale and a mass scale associated with the graviton mass, and $\mathcal{W}$ is the potential term for the metric,
\begin{equation}
\begin{aligned}
 \mathcal{W} \equiv \:  &
 \frac{\tilde{N}}{N}\mathcal{E} + \tilde{\mathcal{E}}
 +  \frac{\tilde{N}\lambda}{N} \left( \hat{\mathcal{F}}^i_{\ j}\tilde{\zeta}^j_{\ i} - \tilde{\mathcal{E}}\tilde{\zeta}^i_{\ i} + \tilde{\mathcal{F}}^i_{\ j}K^{jk}\gamma_{ki}\right)\\
 & + \frac{\tilde{N}}{N}\tilde{\mathcal{F}}^i_{\ j}\mathcal{D}_i\lambda^j
 - \frac{m_g^2 \tilde{N}^2 \lambda^2}{4 N^2} \left(\left[\tilde{\mathcal{F}}^2\right] - \frac{1}{2}\left[\tilde{\mathcal{F}}\right]^2\right)\,.
\end{aligned}
\end{equation}
Here, $\mathcal{D}_p$ is the covariant derivative compatible with $\gamma_{ij}$, $K_{ij} = (\partial_t\gamma_{ij} - \mathcal{D}_i N_j - \mathcal{D}_j N_i)/2N$ is the extrinsic curvature, $\left[\mathcal{A}\right]$ denotes the trace of the matrix $\mathcal{A}$,
\begin{equation}
\begin{aligned}
& \mathcal{E} \equiv \sum^3_{i=0} c_i e_{3-i}(\mathfrak{K})\,,\quad
 \tilde{\mathcal{E}} \equiv \sum^4_{i=1} c_i e_{4-i}(\mathfrak{K})\,,\\
 & \hat{\mathcal{E}} \equiv \sum^4_{i=2} c_i e_{5-i}(\mathfrak{K})\,,\quad
 \tilde{\mathcal{F}}^i_{\ j} \equiv \frac{\delta \tilde{\mathcal{E}}}{\delta \mathfrak{K}^j_{\ i}}\,,
 \quad
\hat{\mathcal{F}}^i_{\ j} \equiv \frac{\delta \hat{\mathcal{E}}}{\delta \mathfrak{K}^j_{\ i}}\,,
\end{aligned}
\end{equation}
$e_i(X)$ ($i=0,1,2,3$) are the three-dimensional symmetric polynomials and $\{c_n\}_{n=0..4}$ are constants. While MTMG is presented here in a compact fashion, a more comprehensive description of its construction and a study of cosmological solutions can be found in \cite{mtmg,mtmg_pheno}. 

\noindent{\bf\em Two branches of solutions.}
Taking FLRW forms for $g_{\mu\nu}$ and $f_{\mu\nu}$ in the unitary gauge, the equation of motion (EOM) for $\lambda$ is factorized as \cite{mtmg_pheno}
\begin{equation}
    \left(c_3 + 2 c_2 \mathcal{X} + c_1 \mathcal{X}^2 \right)\tilde{E} = 0\,,
\end{equation}
where $\mathcal{X}$ ($\ne 0$) is the ratio of the two scale factors and $\tilde{E}$ is some more complicated expression (see Eq.~(83) of \cite{mtmg_pheno}).
This equation reveals the existence of two branches. 

In the so-called ``self-accelerating'' branch, $c_3 + 2 c_2 \mathcal{X} + c_1 \mathcal{X}^2$ vanishes, and thus $\mathcal{X}$ is constant. At cosmological scales, the phenomenology in this branch is the same as in GR, albeit with an effective cosmological constant originating from the graviton mass term, and with a non-zero mass (of order of today's Hubble parameter) given to the gravitational waves. On the other hand, in this letter, we investigate, for the first time, the phenomenology at shorter distances for spherically symmetric non-linear solutions.

The other branch (in which $\tilde{E}$ vanishes), called the ``normal'' branch, has an interesting phenomenology \cite{mtmg_obs, mtmg_isw}, e.g.\ a modified growth of perturbations, but will not be studied here. It is important to note that the graviton mass term can lead to an effective cosmological constant also in this branch. 

\noindent{\bf\em Spatially flat solutions.}
The present work relies upon the
following lemma, which is then used to derive black hole and star solutions.

\emph{Lemma :} Any GR solution that can be written with flat constant-time surfaces is a solution of the self-accelerating branch of MTMG, with the additional feature of a bare cosmological constant.
\begin{proof}
A metric with flat constant-time surfaces is written as 
\begin{equation}
\mathrm{d} s^2 = - \alpha^2\mathrm{d} t^2 + a^2(t)\,\delta_{ij}^S\left(\mathrm{d} x^i + \beta^i \mathrm{d} t\right)\left(\mathrm{d} x^j + \beta^j \mathrm{d} t\right)\,,
\label{eq_line_element_gen}
\end{equation}
where $\alpha(x^\mu)$ and $\beta^i(x^\mu)$ are free functions of the $4$-dimensional coordinates $x^\mu = (t,r,\theta,\varphi)$, $a(t)$ is a function of $t$ corresponding to the scale factor and $\delta_{ij}^S\mathrm{d} x^i \mathrm{d} x^j = \mathrm{d}r^2 + r^2 (\mathrm{d}\theta^2 + \sin^2\theta \mathrm{d}\varphi^2)$. Similarly, a fiducial metric is written as
\begin{equation}
 f_{\mu\nu}\mathrm{d}x^{\mu}\mathrm{d}x^{\nu}
  = - \tilde{N}^2(t) \mathrm{d} t^2 + a^2_f(t)\,\delta_{ij}^S\mathrm{d} x^i \mathrm{d} x^j\,,
\end{equation}
and the Lagrange multipliers are kept in the most general form $\lambda = \lambda(x^\mu)$ and $\lambda^i = \lbrace \lambda^r(x^\mu), \lambda^\theta(x^\mu),\lambda^\varphi(x^\mu) \rbrace$.

The EOM for $\lambda$ reveals the same splitting in two branches as in cosmology, and we will work in the self-accelerating one, where $\mathcal{X} \equiv a_f/a$ is constant. In this branch, the EOM for $\lambda^i$ are automatically satisfied, so one can safely choose the Lagrange multipliers to be equal to their cosmological values, $\lambda = \lambda^i =0$. With this choice, the Einstein equation for the metric (\ref{eq_line_element_gen}) is
\begin{equation}
M_{\rm Pl}^2 \left[ G_{\mu\nu} + \frac{m_g^2}{2} \left(c_4 + 2 c_3 \mathcal{X} + c_2 \mathcal{X}^2 \right) g_{\mu\nu} \right] = T_{\mu\nu}\,.
\end{equation}
Thus the Einstein equation is indeed the same as in GR with an effective cosmological constant
\begin{equation}
\Lambda_{\rm eff}  \equiv \frac{m_g^2}{2} \left(c_4 + 2 c_3 \mathcal{X} + c_2 \mathcal{X}^2 \right).
\label{eq:CCeff}
\end{equation}
\end{proof}

Although this lemma concerns a wide class of solutions, one has either to explicitly write the diffeomorphism to put them in the form of (\ref{eq_line_element_gen}), or to derive a more general geometrical argument to find all GR solutions that permit flat spatial sections. This work adopts the former and demonstrates the existence of black hole and star solutions in MTMG as a corollary of the lemma. Note that while the chosen examples are spherically symmetric systems, this assumption was not made in the lemma.

\noindent{\bf\em Spatially flat slicing.}
In general, one can write the metric for a spherically symmetric system as
\begin{equation}
\mathrm{d} s^2 = - f(t,r)\, \mathrm{d} t^2 +  \frac{\mathrm{d} r^2}{1-\frac{2m(t,r)}{r}} + r^2 \mathrm{d}\Omega^2\,.\label{eq:met_schw_td}
\end{equation}
By a change of coordinates $t\rightarrow\tau + T(\tau,r)$ satisfying
\begin{equation}
\left(\frac{\partial T(\tau,r)}{\partial r}\right)^2 = \frac{1}{f(\tau,r)}\frac{2m(\tau,r)}{r-2m(\tau,r)}\,,\label{eq:cond_td}
\end{equation}
the metric (\ref{eq:met_schw_td}) can be put in a spatially flat form
\begin{equation}
\mathrm{d} s^2 = - N^2\mathrm{d}\tau^2 + \left[\left( \mathrm{d} r + \beta\, \mathrm{d} \tau \right)^2 + r^2 \mathrm{d}\Omega^2\right]\,,\label{eq:met_pg_td}
\end{equation}
with
\begin{equation}
N^2= (1+\dot{T})^2 f (1+T'^2 f) \,,\quad \beta = -T' (1 + \dot{T})f\,,
\label{eq:cond_Nbeta}
\end{equation}
where a dot and a prime denote time and radial derivatives, respectively. 

\noindent{\bf\em Static solutions in vacuum.}
The Schwarzschild-de-Sitter solution is given by $m(t,r) = m(r) = M - \Lambda r^3/6$ with both $M$ and $\Lambda$ constant and $f(r,t) = f(r) = 1 - 2 m(r)/r$. Applying the transformation as in Eq.~(\ref{eq:cond_Nbeta}), one finds that
\begin{equation}
 \mathrm{d} s^2 = -\mathrm{d}\tau^2 + \left(\mathrm{d}r \pm \sqrt{\frac{2M}{r} - \frac{\Lambda r^2}{3}} \mathrm{d}\tau\right)^2
  + r^2\mathrm{d}\Omega^2\,.\label{eq:met_pg_schwdS}
\end{equation}
Going to the pure Schwarzschild solution (\emph{i.e.}\ taking $\Lambda =0$), this particular form of the metric has been long known and was firstly independently proposed by Painlev\'{e} \cite{Painleve} and Gullstrand \cite{Gull}. 

By the lemma, the metric (\ref{eq:met_pg_schwdS}) is a solution of MTMG for a flat fiducial metric $\tilde{\gamma}_{ij} = \mathcal{X}^2 \delta_{ij}^S$ with $\mathcal{X}$ constant and $\lambda = \lambda^i = 0$~\footnote{More generally, \emph{i.e.}\ beyond the lemma, $\lambda(r) = \lambda_0$ and $\lambda^i\partial_i = \mp \lambda_0 \sqrt{2M/r-\Lambda r^2/3}\,\partial_r$ are also allowed, where $\lambda_0$ is a constant. However, in order to match with their cosmological boundary condition, one has to impose $\lambda_0 =0$ so that $\lambda = \lambda^i =0$.}, provided that $\Lambda=\Lambda_{\rm eff}$. This demonstrates the existence of static black holes in MTMG, identical to those in GR. 

The solution (\ref{eq:met_pg_schwdS}) can also be used to describe the outer part of a non-rotating star. Thus the PPN parameters $\beta^{\rm PPN}$ and $\gamma^{\rm PPN}$ are unity, as in GR. In the following we turn to its inner part, \emph{i.e.}\ the inclusion of matter in MTMG.

\noindent{\bf\em Inclusion of matter.} Since in MTMG (and more generally in massive gravity) diffeomorphisms are broken by the graviton mass term (in the unitary gauge), solutions equivalent under a change of coordinates in GR become different solutions in MTMG. It is therefore important to make sure that there is no coordinate singularity at the center of solutions with matter. 

As a condition we require that in the $r \to 0$ limit, the extrinsic curvature remains regular and becomes isotropic, and thus that the anisotropic part of the extrinsic curvature, $K_{rr}-K/3$, where $K$ is the trace of the extrinsic curvature, vanishes at the center. For the metric ansatz (\ref{eq:met_pg_td}) it is sufficient to show that
\begin{equation}
\lim_{r\rightarrow0} \left(\partial_r\beta-\frac{\beta}{r}\right) = 0\,.\label{eq:condition_beta}
\end{equation}
For this purpose we expand all quantities around $r=0$, for example
$
\beta(\tau, r) = \sum_{n = 0}^\infty \beta_n(\tau) r^n
$.
In the following equations the $n$-th subscript will denote the coefficient of $n$-th power of $r$ in the corresponding quantity. The following argument does not depend on the effective equation of state and applies to both dynamical and static configurations. In the lowest order in $r$, the $\tau\tau$ component of the Einstein equation becomes $M_{\rm Pl}^2\beta_0^2/r^2 = 0$. This imposes that $\beta_0 = 0$. Then, taking the next relevant order in $r$, one finds that the $\tau\tau$ and $rr$ components of the Einstein equation are $N_0^2 \rho_0 = 3M_{\rm Pl}^2\beta_1^2/2$ and $N_1/N_0 = 0$, leaving $N_0 \neq 0$ and $N_1 = 0$. Iterating this procedure yields $\beta_2 = 0$ and thus 
\begin{equation}
\beta(\tau,r) = \pm\frac{N_0}{M_{\rm Pl}} \sqrt{\frac{2}{3}\rho_0}\, r + \mathcal{O}(r^3)\,,\label{eq:beta}
\end{equation}
in which we have denoted the matter density of the fluid by $\rho$, assumed a general barotropic equation of state $P=P(\rho)$, and omitted here the negligible contribution from the effective cosmological constant term.

Regular solutions found here include static solutions with matter as a special case. As a concrete and simple example let us consider the interior Schwarzschild solution, as described for instance in \cite{stephani}. The solution, after matching to the Schwarzschild solution with Schwarzschild radius $2M$ at the stellar radius $r_0$, has
\begin{equation}
 m(r) = M\, \frac{r^3}{r_0^3}\,,\
  f = \left[\frac{3}{2}\sqrt{1-\frac{2 M}{r_0}}-\frac{1}{2}\sqrt{1-\frac{2m(r)}{r}}\right]^{\!2},
\end{equation}
where we have once again omitted the contribution of the effective cosmological constant as physically negligible at scales of order of the stellar radius. After transformation (\ref{eq:cond_td}), we recover a spatially flat space-time with
\begin{equation}
N^2= \frac{rf}{r-2m} \,,\quad \beta = \pm \sqrt{\frac{2mf}{r-2m}}\,.
\end{equation}
As seen in the lemma this is a solution of MTMG with $\lambda = \lambda^i = 0$.

The presence of time-dependent solutions with matter allows one to construct non-homogeneous cosmological solutions, as discussed in \cite{dRGTcosmo} for dRGT gravity. Those solutions are good approximations of the canonical FLRW universe for patches of the sizes larger than the Vainshtein radius. In MTMG we can go beyond this as we shall see in the following. 

\noindent{\bf\em Matching to cosmology.}
The special form of the metric (\ref{eq_line_element_gen}) allows for a non-trivial scale factor $a(t)$, which has not yet been discussed in the examples so far.
To include non-trivial $a(t)$, one shall also implement a non-trivial scale factor in the fiducial sector as $\tilde{\gamma}_{ij} = a_f^2(t) \delta_{ij}^S$, so that the ratio $\mathcal{X} \equiv a_f /a$ can be constant for an expanding universe ($\dot{a}>0$), as required in the self-accelerating branch. 
In MTMG this class of solutions is distinct from the solution (\ref{eq:met_pg_schwdS}). Indeed, as exposed previously, solutions equivalent under a change of coordinates in GR become different solutions in MTMG in the unitary gauge and thus the Schwarzschild-de-Sitter black hole solution of the form (\ref{eq:met_pg_schwdS}) is not equivalent to a black hole embedded in an expanding universe with an exponential scale factor. 

Beginning with pure GR and the line-element (\ref{eq:met_schw_td}) with $f(t,r) = 1 - 2m(r)/r = 1-2M/r + \Lambda r^2/3$ and both $M$ and $\Lambda$ constant, the aim is to write it in a ``generalized Painlev\'{e}-Gullstrand'' form
\begin{equation}
    \mathrm{d} s^2 = - N(r,t)^2 \mathrm{d} t^2 + a^2(t) \left[ \left(\mathrm{d} r + \beta(r,t) \, \mathrm{d} t \right)^2 + r^2 \mathrm{d} ^2 \Omega \right].
    \label{eq:gen_PG}
\end{equation}
This can be done with an appropriate coordinate change of the form (\ref{eq:cond_td}) together with $r \rightarrow a(t) r$, to have
\begin{equation}
    N(r,t) = 1, \qquad 
    \beta(r,t) = \frac{\dot{a}}{a} r  \pm \sqrt{\frac{2 M}{a^3 r} - \frac{\Lambda \, r^2}{3}}\,,
    \label{eq:PG_gen_GR}
\end{equation}
with $a(t)$ being unspecified.

Turning to MTMG, we inject the metric (\ref{eq:gen_PG}) in the EOM with a spherically symmetric ansatz for the Lagrange multipliers as $\lambda = \lambda(r,t)$ and $\lambda^i\partial_i = \lambda^r(r,t)\partial_r$ This yields
\begin{equation}
     N(r,t) = 1, \qquad
      \beta(r,t) = \frac{\dot{a}}{a} r  \pm \sqrt{\frac{2 \mu(t)}{a^3r} - \frac{\Lambda_{\rm eff} \, r^2}{3}}\,,
      \label{eq:PG_gen_MTMG}
\end{equation}
with the modified mass
\begin{equation}
    \mu(t) = M_0 \pm \frac{m_g^2\left(c_1 \mathcal{X} + c_2 \right) \mathcal{X} }{2\sqrt{3}} \int_{-\infty}^t \!\! \mathrm{d} \tau \, a^3(\tau)\tilde{N}(\tau)\tilde{\lambda}(\tau)\,,
    \label{eq:mod_mass}
\end{equation}
where $M_0$ is a constant and the sign $\pm$ is consistent with Eq.~(\ref{eq:beta}). The Lagrange multipliers $\lambda$ and $\lambda^r$ are 
\begin{equation}
    \lambda(r,t) = 0, \qquad
    \lambda^r(r,t) = \frac{\tilde{\lambda}(t)}{\sqrt{r\left(\sqrt3 \mu(t) - \Lambda_{\rm eff} \, r^3\right)}}\,,      \label{eq:PG_gen_MTMG_lambdas}
\end{equation}
where $\tilde{\lambda}(t)$ is an arbitrary function of $t$. 

Considering a BH or an exterior solution of a star formed from smooth and asymptotically FLRW initial data with matter, both $\lambda$ and $\lambda^r$ should vanish at spatial infinity to recover their cosmological values. Also, $\lambda^r(t,r)$ should vanish at $r=0$ for regularity, at least until a physical singularity forms there. Moreover, $\lambda=0$ everywhere for the final configuration as shown in (\ref{eq:PG_gen_MTMG_lambdas}), and finally the Lemma tells us that $\lambda=\lambda^r=0$ is a solution all the time from the initial data to the final configuration. By continuity we thus conjecture that $\tilde{\lambda}(t)=0$ in (\ref{eq:PG_gen_MTMG_lambdas}), meaning that $\mu(t)=M_0$ in (\ref{eq:mod_mass}). In this case the solution (\ref{eq:PG_gen_MTMG}) in MTMG recovers the solution (\ref{eq:PG_gen_GR}) in GR, with $M_0=M$ and $\Lambda_{\rm eff}=\Lambda$. 

Contrary to the solutions found in \cite{dRGTcosmo} and \cite{Gumrukcuoglu:2011ew} for dRGT gravity, this solution (with $\tilde{\lambda}=0$) in the $M_0\to 0$ limit is strictly homogeneous and isotropic and is free from strong coupling issues, which seems to be a unique and significant feature of MTMG among massive gravity theories. 

\noindent{\bf\em Discussion.}
In previous works, the minimal theory of massive gravity (MTMG) has proven to successfully pass cosmological consistency tests \cite{mtmg_pheno,mtmg_obs,mtmg_isw}. Here, for the first time, we have studied the existence of black hole and star solutions in the context of MTMG. The main result of this paper is the lemma by which, for the class of space-times that can be put into a spatially flat form by an appropriate change of coordinates, any GR solutions are also solutions of the self-accelerating branch of MTMG. 

In order to illustrate the lemma, we have further presented a collection of corollaries: (i) spherically symmetric static solutions in vacuum with or without cosmological constant, (ii) spherically symmetric solutions with matter which are either time dependent or time independent, and (iii) a Schwarzschild-de-Sitter solution matched with a de Sitter background in the FLRW form. These examples are of course not expected to be exhaustive for the class of space-times concerned by the lemma. Arguably, other types of GR solutions may also find a corresponding MTMG solution; this is left as an interesting point for future study. Some particular settings would also require further attention, due to their physical relevance. For example it would be important to study Kerr-like rotating systems. Of other particular interest would be the study of isolated objects matched to different cosmological backgrounds, e.g.\ matter and/or radiation dominated universes. 

Due to the breaking of general covariance by the mass term, coordinate transformations in the unitary gauge are not innocuous. In particular, spatially flat coordinates cannot be changed back to the usual Schwarzschild-like coordinates unless St\"{u}ckelberg fields are introduced. This is a fundamental change with respect to GR. It is in particular for this reason that matching with background cosmological solutions is unavoidable in MTMG. As emphasized above, we expect that this can be done in more generality than what was presented in the present work.

To our knowledge, the existence of static, black hole configurations without strong coupling in massive gravity is specific to MTMG. Indeed, it has been shown that in dRGT gravity, no static solution is healthy, although there exist time-dependent non-GR solutions \cite{RosenBH}. MTMG solutions identically match GR solutions, there are no singularities except for those already existing in GR and they are free from strong couplings. We leave for the future the exploration of the stability of the solutions found in this work, although it seems reasonable to expect the same stability properties as the GR solutions.

The existence of the Schwarzschild solution in MTMG has some direct consequences, in particular the values of some PPN parameters. One has that $\gamma^{\rm PPN} = \beta^{\rm PPN} = 1$, values that are the same as in GR. A more detailed study would be necessary to obtain all other PPN parameters.

The propagation of GW in MTMG is different from GR but this difference is negligible since the sound speed of GW in the subhorizon limit is equal to the speed of light \cite{mtmg, mtmg_pheno} and the expected mass of GW in MTMG, of order $H_0$, is still well below the current upper bounds from observation. On the other hand, the production of GW from binary systems may exhibit observable differences from GR and thus it is interesting to investigate the production process of GW in the context of MTMG.

Finally, we will leave for future studies the alternative branch of MTMG, known as the ``normal'' branch. While the self-accelerating branch is the closest to GR, the phenomenology of the normal branch may allow for interesting deviations, such as the ones in \cite{mtmg_pheno, mtmg_obs, mtmg_isw}. 

\noindent{\bf\em Acknowledgements.} ADF was supported by JSPS KAKENHI Grant No.\ 16K05348. FL would like to express all his gratitude to YITP, Kyoto U for hosting him. The work of SM was supported by JSPS KAKENHI No.\ 17H02890, No.\ 17H06359, and by WPI, MEXT, Japan.  MO acknowledges the support from the Japanese Government (MEXT) Scholarship for Research Students.

\end{document}